\documentclass{article}%
\usepackage{amsmath}%
\usepackage{amsfonts}%
\usepackage{amssymb}%
\usepackage{graphicx}
\newtheorem{theorem}{Theorem}

\newtheorem{definition}[theorem]{Definition}

\newtheorem{proposition}[theorem]{Proposition}

\newenvironment{proof}[1][Proof]{\noindent\textbf{#1.} }{\ \rule{0.5em}{0.5em}}
\begin{document}

\title{Quantum Computing, Postselection, and Probabilistic Polynomial-Time}
\author{Scott Aaronson\footnote{Part of this work was done while I was a graduate student at
the University of California, Berkeley, CA (USA), supported by an
NSF Graduate Fellowship.}\\Institute for Advanced Study, Princeton,
NJ (USA)\\aaronson@ias.edu}
\date{}
\maketitle

\begin{abstract}
I study the class of problems efficiently solvable by a quantum
computer, given the ability to \textquotedblleft
postselect\textquotedblright\ on the outcomes of measurements. \ I
prove that this class coincides with a classical complexity class
called $\mathsf{PP}$, or Probabilistic Polynomial-Time.\ \ Using
this result, I show that several simple changes to the axioms of
quantum mechanics would let us solve $\mathsf{PP}$-complete problems
efficiently. \ The result also implies, as an easy corollary, a
celebrated theorem of Beigel, Reingold, and Spielman that
$\mathsf{PP}$\ is closed under intersection, as well as a
generalization of that theorem due to Fortnow and Reingold. \ This
illustrates that quantum computing can yield new and simpler proofs
of major results about classical computation.

\end{abstract}

\section{Introduction\label{INTRO}}

\textit{Postselection} is the power of discarding all runs of a computation in
which a given event does not occur. \ To illustrate, suppose we are given a
Boolean formula in a large number of variables, and we wish to find a setting
of the variables that makes the formula\ true. \ Provided such a setting
exists, this problem is easy to solve using postselection: we simply set the
variables randomly, then postselect on the formula being true.

This paper studies the power of postselection in a quantum computing
context. \ I define a new complexity class called $\mathsf{PostBQP}$
(Postselected Bounded-Error Quantum Polynomial-Time),\ which
consists of all problems solvable by a quantum computer in
polynomial time, given the ability to postselect on a measurement
yielding a specific outcome. \ The main result is that
$\mathsf{PostBQP}$\ equals the well-known classical complexity class
$\mathsf{PP}$ (Probabilistic Polynomial-Time). \ Here $\mathsf{PP}$\
is the class of problems for which there exists a probabilistic
polynomial-time Turing machine that accepts with probability at
least $1/2$ if and only if the
answer is `yes.' \ For example, given a Boolean formula, a $\mathsf{PP}%
$\ machine can decide whether the \textit{majority} of settings to
the variables make the formula true. \ Indeed, this problem turns
out to be $\mathsf{PP}$-complete\ (that is, among the hardest
problems in $\mathsf{PP}$).\footnote{See the ``Complexity Zoo''
(www.complexityzoo.com) for more information about the complexity
classes mentioned in this paper.}

The motivation for the $\mathsf{PostBQP}=\mathsf{PP}$ result comes
from two quite different sources. \ The original motivation was to
analyze the computational power of \textquotedblleft
fantasy\textquotedblright\ versions of quantum mechanics, and
thereby gain insight into why quantum mechanics is the way it is. \
In particular, Section \ref{FANTASY} will show that if we changed
the measurement probability rule from $\left\vert \psi\right\vert
^{2}$\ to $\left\vert \psi\right\vert ^{p}$\ for some $p\neq2$, or
allowed linear but nonunitary evolution, then we could simulate
postselection, and thereby solve $\mathsf{PP}$-complete problems in
polynomial\ time. \ If we consider such an ability extravagant, then
we might take these results as helping to explain why quantum
mechanics is unitary, and why the measurement rule is $\left\vert
\psi\right\vert ^{2}$.

A related motivation comes from an idea that might be called \textit{anthropic
computing}---arranging things so that we are more likely to exist if a
computer produces a desired output than if it does not. \ As a simple example,
under the many-worlds interpretation of quantum mechanics, we might kill
ourselves in all universes where a computer fails! \ My result implies that,
using this \textquotedblleft technique,\textquotedblright\ we could solve not
only $\mathsf{NP}$-complete problems efficiently,\ but $\mathsf{PP}%
$-complete\ problems as well.

However, the $\mathsf{PostBQP}=\mathsf{PP}$\ result also has a more
unexpected implication. \ One reason to study quantum computing is
to gain a new, more general perspective on \textit{classical}
computer science. \ By analogy, many famous results in computer
science involve only deterministic computation, yet it is hard to
imagine how anyone could have proved these results had researchers
not long ago \textquotedblleft taken aboard\textquotedblright\ the
notion of randomness.\footnote{A few examples are primality testing
in deterministic polynomial time \cite{aks}, undirected graph
connectivity in log-space \cite{reingold}, and inapproximability of
the 3-SAT problem\ unless $\mathsf{P}=\mathsf{NP}$ \cite{hastad}.} \
Likewise, taking quantum mechanics aboard has already led to some
new results about classical computation \cite{aar:pls,ar,kw,regev}.
\ What this paper will show is that, even when classical results are
already known, quantum computing can sometimes provide new and
simpler proofs for them.

When Gill \cite{gill:thesis,gill}\ defined $\mathsf{PP}$\ in 1972, he also
asked a notorious question that remained open for eighteen years: is
$\mathsf{PP}$\ closed under intersection?\footnote{It is clear that
$\mathsf{PP}$\ is closed under complement, so this question is equivalent to
asking whether $\mathsf{PP}$\ is closed under union.} \ In other words, given
two probabilistic polynomial-time Turing machines $A$ and $B$, does there
exist another such machine that accepts with probability greater
than\ $1/2$\ if and only if $A$ and $B$ both do? \ The question was finally
answered in the affirmative by Beigel, Reingold, and Spielman \cite{brs}, who
introduced a brilliant technique for representing the logical AND of two
majority functions by the sign of a low-degree rational function. \ Fortnow
and Reingold \cite{fr}\ later extended the technique to show that
$\mathsf{PP}$\ is closed under \textquotedblleft polynomial-time truth-table
reductions.\textquotedblright\ \ This means that a polynomial-time
Turing\ machine that makes nonadaptive queries to a $\mathsf{PP}$\ oracle is
no more powerful than $\mathsf{PP}$\ itself.\footnote{A query is
\textquotedblleft nonadaptive\textquotedblright\ if it does not depend on the
answers to previous queries. \ Beigel \cite{beigel}\ has given evidence that
the Fortnow-Reingold result does not generalize to adaptive queries.}

Now the class $\mathsf{PostBQP}$\ is trivially closed under
intersection, as well as under polynomial-time truth-table
reductions. \ So the fact that $\mathsf{PostBQP}=\mathsf{PP}$\
immediately implies the Beigel et al. \cite{beigel}\ and
Fortnow-Reingold \cite{fr}\ results. \ Indeed, it even implies that
$\mathsf{PP}$\ is closed under \textit{quantum} polynomial-time
truth-table reductions, which seems to be a new result. \ I should
emphasize that the $\mathsf{PostBQP}=\mathsf{PP}$ proof is about one
page long, and does not use rational functions or any other
heavy-duty mathematics.\footnote{Although, as pointed out to me by
Richard Beigel, a result on rational functions similar to those in
\cite{brs} could be \textit{extracted} from my result.}

This paper is based on chapter 15 of my PhD thesis
\cite{aar:thesis}. \ Some of the results appeared in preliminary
form in \cite{aar:isl}, before I made the connection to showing
$\mathsf{PP}$ closed under intersection.

\section{Related Work\label{RELATED}}

Besides $\mathsf{PostBQP}$, several other \textquotedblleft
nondeterministic\textquotedblright\ versions of $\mathsf{BQP}$ have appeared
in the literature. \ Adleman, DeMarrais, and Huang \cite{adh}\ defined
$\mathsf{NQP}$\ to be the class of problems for which there exists a
polynomial-time quantum algorithm that accepts with nonzero probability if and
only if the answer is `yes.' \ Then, in a result reminiscent of this paper's,
Fenner et al. \cite{fghp} showed that $\mathsf{NQP}$\ equals a classical class
called $\mathsf{coC}_{\mathsf{=}}\mathsf{P}$.\footnote{Unfortunately, I do not
know of any classical result that this helps to prove. \ That $\mathsf{coC}%
_{\mathsf{=}}\mathsf{P}$\ is closed under union and intersection is obvious.}
\ Also, Watrous \cite{watrous}\ defined $\mathsf{QMA}$\ as the class of
problems for which a polynomial-time quantum verifier can be convinced of a
`yes' answer by a polynomial-size quantum proof. \ If we require the proof to
be classical, then we obtain the apparently weaker class $\mathsf{QCMA}$,
defined by Aharonov and Naveh \cite{an}. \ Note that all of these classes are
contained in $\mathsf{PP}$, and hence in $\mathsf{PostBQP}$.

The idea of postselection has recurred several times in quantum computing.
\ For example, Terhal and DiVincenzo \cite{td} used postselection to show that
constant-depth quantum circuits are probably hard to simulate, and I
\cite{aar:adv}\ used it to show that $\mathsf{BQP/qpoly}\subseteq
\mathsf{PP/poly}$\ (that is, any problem solvable in $\mathsf{BQP}$\ with
polynomial-size quantum advice, is also solvable in $\mathsf{PP}$\ with
polynomial-size classical advice).

If we add postselection to a \textit{classical} probabilistic polynomial-time
Turing machine, then we obtain a complexity class known as $\mathsf{BPP}%
_{\mathsf{path}}$, which was defined by Han, Hemaspaandra, and Thierauf
\cite{hht} and which sits somewhere between $\mathsf{MA}$ and $\mathsf{PP}$
($\mathsf{MA}$\ being a probabilistic generalization of $\mathsf{NP}$).

Fortnow \cite{fortnow:blog}\ reports that in 1990, he, Fenner, and Kurtz tried
to show $\mathsf{PP}$\ closed under intersection by (1) defining a seemingly
weaker class, (2) showing that class closed under intersection, and then (3)
showing that it actually equals $\mathsf{PP}$. \ The attempt failed, and soon
thereafter Beigel et al. \cite{beigel}\ succeeded with a quite different
approach. \ This paper could be seen as a vindication of Fortnow et al.'s
original approach---all it was missing was quantum mechanics! \ Admittedly, Li
\cite{li} gave another proof along Fortnow et al.'s lines in 1993. \ However,
Li's proof makes heavy use of rational functions, and seems less intuitive
than the one here.

\section{The Class $\mathsf{PostBQP}$\label{POSTBQP}}

In what follows, I assume basic familiarity with quantum computing, and in
particular with the class $\mathsf{BQP}$\ (Bounded-Error Quantum Polynomial
Time) defined by Bernstein and Vazirani \cite{bv}. \ It is now time to define
$\mathsf{PostBQP}$\ more formally.

\begin{definition}
\label{postbqpdef}$\mathsf{PostBQP}$ is the class of languages $L\subseteq
\left\{  0,1\right\}  ^{\ast}$ for which there exists a uniform\footnote{Here
`uniform' means that there exists a classical algorithm that outputs a
description of $C_{n}$\ in time polynomial in $n$.} family of polynomial-size
quantum circuits $\left\{  C_{n}\right\}  _{n\geq1}$\ such that for all inputs
$x$,

\begin{enumerate}
\item[(i)] After $C_{n}$ is applied to the state $\left\vert 0\cdots
0\right\rangle \otimes\left\vert x\right\rangle $, the first qubit has a
nonzero probability of being measured to be $\left\vert 1\right\rangle $.

\item[(ii)] If $x\in L$, then conditioned on the first qubit being $\left\vert
1\right\rangle $, the second qubit is $\left\vert 1\right\rangle $\ with
probability at least $2/3$.

\item[(iii)] If $x\notin L$, then conditioned on the first qubit being
$\left\vert 1\right\rangle $, the second qubit is $\left\vert 1\right\rangle
$\ with probability at most $1/3$.
\end{enumerate}
\end{definition}

It is immediate that $\mathsf{NP}\subseteq\mathsf{PostBQP}$. \ Also, to show
$\mathsf{PostBQP}\subseteq\mathsf{PP}$, we can use the same observations used
by Adleman, DeMarrais, and Huang \cite{adh}\ to show that $\mathsf{BQP}%
\subseteq\mathsf{PP}$, but sum only over paths where the first qubit is
$\left\vert 1\right\rangle $\ at the end. \ In more detail:

\begin{proposition}
\label{adhgen}$\mathsf{PostBQP}\subseteq\mathsf{PP}$.
\end{proposition}

\begin{proof}
By a result of Shi \cite{shi:gate}, we can assume without loss of generality
that our quantum circuit is composed of Hadamard and Toffoli
gates.\footnote{This is true even for a \textit{postselected} quantum circuit,
since by the Solovay-Kitaev Theorem \cite{kitaev:ec}, we can achieve the
needed accuracy in amplitudes at the cost of a polynomial increase in circuit
size.} \ Then the final amplitude $\alpha_{z}$\ of each basis state
$\left\vert z\right\rangle $\ can be written as a sum of exponentially many
contributions, call them $a_{z,1},\ldots,a_{z,N}$, each of which is a rational
real number computable in classical polynomial time. \ So the final
probability of $\left\vert z\right\rangle $ equals%
\[
\alpha_{z}^{2}=\left(  a_{z,1}+\cdots+a_{z,n}\right)  ^{2}=\sum_{ij}%
a_{z,i}a_{z,j}.
\]
We need to test which is greater: the sum $S_{0}$\ of $\alpha_{z}^{2}$\ over
all $z$ beginning with $10$, or the sum $S_{1}$\ of $\alpha_{z}^{2}$\ over all
$z$ beginning with $11$. \ But we can do this in $\mathsf{PP}$: we simply put
the positive contributions $a_{z,i}a_{z,j}$\ to $S_{1}$ and negative
contributions to $S_{0}$ on \textquotedblleft one side of the
ledger,\textquotedblright\ and the negative contributions to $S_{1}$\ and
positive contributions to $S_{0}$\ on the other side.
\end{proof}

How robust is $\mathsf{PostBQP}$? \ Just as Bernstein and Vazirani
\cite{bv}\ showed that intermediate measurements do not increase the power of
ordinary quantum computers, so it is easily shown that intermediate
postselection steps do not increase the power of $\mathsf{PostBQP}$.
\ Whenever we want to postselect on a qubit $j$ being $\left\vert
1\right\rangle $, we simply apply a CNOT gate from $j$ into a fresh ancilla
qubit that is initialized to $\left\vert 0\right\rangle $ and that will never
be written to again. \ Then, at the end, we compute the AND of the ancilla
qubits, and swap the result into the first qubit. \ By a standard Chernoff
bound, it follows that we can repeat a $\mathsf{PostBQP}$\ computation a
polynomial number of times, and thereby reduce the error probability from
$1/3$\ to $1-2^{-p\left(  n\right)  }$\ for any polynomial $p$.

A corollary of the above observations is that $\mathsf{PostBQP}$\ has strong
closure properties.

\begin{proposition}
\label{closure}$\mathsf{PostBQP}$ is closed under union, intersection, and
complement. \ Indeed, it is closed under $\mathsf{BQP}$\ truth-table
reductions, meaning that $\mathsf{PostBQP}=\mathsf{BQP}_{\mathsf{\Vert
,}\operatorname*{classical}}^{\mathsf{PostBQP}}$,\ where $\mathsf{BQP}%
_{\mathsf{\Vert,}\operatorname*{classical}}^{\mathsf{PostBQP}}$\ is the class
of problems solvable by a $\mathsf{BQP}$\ machine\ that can make a polynomial
number of nonadaptive classical queries to a $\mathsf{PostBQP}$ oracle.
\end{proposition}

\begin{proof}
Clearly $\mathsf{PostBQP}$ is closed under complement, since the definition is
symmetric with respect to $x\in L$\ and $x\notin L$. \ For closure under
intersection,\ let $L_{1},L_{2}\in\mathsf{PostBQP}$;\ then we need to decide
whether $x\in L_{1}\cap L_{2}$. \ Run amplified\ computations (with error
probability at most $1/6$) to decide if $x\in L_{1}$\ and if $x\in L_{2}$,
postselect on both computations succeeding, and accept if and only if both
accept. \ It follows that $\mathsf{PostBQP}$\ is closed under union as well.

In general, suppose a $\mathsf{BQP}_{\mathsf{\Vert,}\operatorname*{classical}%
}^{\mathsf{PostBQP}}$\ machine $M$\ submits queries $q_{1},\ldots,q_{p\left(
n\right)  }$\ to the $\mathsf{PostBQP}$\ oracle. \ Then run amplified
computations (with error probability at most, say, $\frac{1}{10p\left(
n\right)  }$) to decide the answers to these queries, and postselect on all
$p\left(  n\right)  $ of them succeeding. \ By the union bound, if $M$ had
error probability $\varepsilon$\ with a perfect $\mathsf{PostBQP}$\ oracle,
then its new error probability is at most $\varepsilon+1/10$, which can easily
be reduced through amplification.
\end{proof}

One might wonder why Proposition \ref{closure}\ does not go through with
\textit{adaptive} queries. \ The reason is subtle: suppose we have two
$\mathsf{PostBQP}$\ computations, the second of which relies on the output of
the first. \ Then even if the first computation is amplified a polynomial
number of times, it still has an exponentially small probability of error.
\ But since the second computation uses postselection, \textit{any} nonzero
error probability could be magnified arbitrarily, and is therefore too large.

I now prove the main result.

\begin{theorem}
\label{postbqppp}$\mathsf{PostBQP}=\mathsf{PP}$.
\end{theorem}

\begin{proof}
We have already observed that $\mathsf{PostBQP}\subseteq\mathsf{PP}$. \ For
the other direction, let $f:\left\{  0,1\right\}  ^{n}\rightarrow\left\{
0,1\right\}  $\ be an efficiently computable Boolean function, and let
$s=\left\vert \left\{  x:f\left(  x\right)  =1\right\}  \right\vert $. \ Then
we need to decide in $\mathsf{PostBQP}$\ whether $s<2^{n-1}$ or $s\geq2^{n-1}%
$. \ (As a technicality, we can guarantee using padding that $s>0$.)

The algorithm is as follows: first prepare the state $2^{-n/2}\sum
_{x\in\left\{  0,1\right\}  ^{n}}\left\vert x\right\rangle \left\vert f\left(
x\right)  \right\rangle $. \ Then following Abrams and Lloyd \cite{al}, apply
Hadamard gates to all $n$ qubits in the first register and
postselect\footnote{Postselection is actually overkill here, since the first
register has at least $1/4$ probability of being $\left\vert 0\right\rangle
^{\otimes n}$.} on that register being $\left\vert 0\right\rangle ^{\otimes
n}$.\ \ This produces the state $\left\vert 0\right\rangle ^{\otimes
n}\left\vert \psi\right\rangle $\ where%
\[
\left\vert \psi\right\rangle =\frac{\left(  2^{n}-s\right)  \left\vert
0\right\rangle +s\left\vert 1\right\rangle }{\sqrt{\left(  2^{n}-s\right)
^{2}+s^{2}}}.
\]
Next, for some positive real numbers $\alpha,\beta$\ to be specified later,
prepare $\alpha\left\vert 0\right\rangle \left\vert \psi\right\rangle
+\beta\left\vert 1\right\rangle H\left\vert \psi\right\rangle $ where%
\[
H\left\vert \psi\right\rangle =\frac{\sqrt{1/2}\left(  2^{n}\right)
\left\vert 0\right\rangle +\sqrt{1/2}\left(  2^{n}-2s\right)  \left\vert
1\right\rangle }{\sqrt{\left(  2^{n}-s\right)  ^{2}+s^{2}}}%
\]
is the result of applying a Hadamard gate to $\left\vert \psi\right\rangle $.
\ Then postselect on the second qubit being $\left\vert 1\right\rangle
$.\ This yields the reduced state%
\[
\left\vert \varphi_{\beta/\alpha}\right\rangle =\frac{\alpha s\left\vert
0\right\rangle +\beta\sqrt{1/2}\left(  2^{n}-2s\right)  \left\vert
1\right\rangle }{\sqrt{\alpha^{2}s^{2}+\left(  \beta^{2}/2\right)  \left(
2^{n}-2s\right)  ^{2}}}%
\]
in the first qubit.

Suppose $s<2^{n-1}$, so that $s$ and $\sqrt{1/2}\left(  2^{n}-2s\right)
$\ are both at least $1$. \ Then we claim there exists an integer $i\in\left[
-n,n\right]  $\ such that, if we set $\beta/\alpha=2^{i}$, then $\left\vert
\varphi_{2^{i}}\right\rangle $\ is close to the state $\left\vert
+\right\rangle =\left(  \left\vert 0\right\rangle +\left\vert 1\right\rangle
\right)  /\sqrt{2}$:%
\[
\left\vert \left\langle +|\varphi_{2^{i}}\right\rangle \right\vert \geq
\frac{1+\sqrt{2}}{\sqrt{6}}>0.985.
\]
For since $\sqrt{1/2}\left(  2^{n}-2s\right)  /s$ lies between $2^{-n}$\ and
$2^{n}$, there must be an integer $i\in\left[  -n,n-1\right]  $ such that
$\left\vert \varphi_{2^{i}}\right\rangle $\ and $\left\vert \varphi_{2^{i+1}%
}\right\rangle $\ fall on opposite sides of $\left\vert +\right\rangle $\ in
the first quadrant (see Figure \ref{ppfig}).\ \ So the worst case is that
$\left\langle +|\varphi_{2^{i}}\right\rangle =\left\langle +|\varphi_{2^{i+1}%
}\right\rangle $, which occurs when $\left\vert \varphi_{2^{i}}\right\rangle
=\sqrt{2/3}\left\vert 0\right\rangle +\sqrt{1/3}\left\vert 0\right\rangle
$\ and $\left\vert \varphi_{2^{i+1}}\right\rangle =\sqrt{1/3}\left\vert
0\right\rangle +\sqrt{2/3}\left\vert 0\right\rangle $. \ On the other hand,
suppose $s\geq2^{n-1}$, so\ that $\sqrt{1/2}\left(  2^{n}-2s\right)  \leq0$.
\ Then $\left\vert \varphi_{2^{i}}\right\rangle $\ never lies in the first or
third quadrants, and therefore $\left\vert \left\langle +|\varphi_{2^{i}%
}\right\rangle \right\vert \leq1/\sqrt{2}<0.985$.%
\begin{figure}
[ptb]
\begin{center}
\includegraphics[
trim=1.7in 3in 1.7in 2.770122in, height=2.3869in, width=2.3694in
]%
{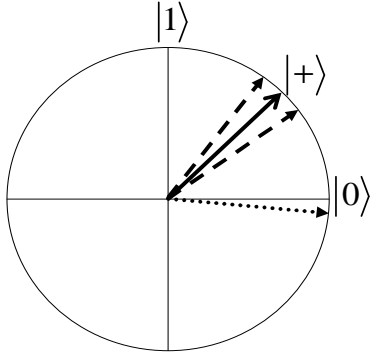}%
\caption{If $s$\ and $2^{n}-2s$\ are both positive, then as we vary the ratio
of $\beta$\ to $\alpha$, we eventually get close to $\left\vert +\right\rangle
=\left(  \left\vert 0\right\rangle +\left\vert 1\right\rangle \right)
/\sqrt{2}$ (dashed lines). \ On the other hand, if $2^{n}-2s$ is not positive
(dotted line), then we never even get into the first quadrant.}%
\label{ppfig}%
\end{center}
\end{figure}

It follows that, by repeating the whole algorithm $n\left(  2n+1\right)
$\ times (as in Proposition \ref{closure}), with $n$\ invocations for each
integer $i\in\left[  -n,n\right]  $, we can learn whether $s<2^{n-1}$ or
$s\geq2^{n-1}$\ with exponentially small probability of error.
\end{proof}

Combining Proposition \ref{closure} with Theorem \ref{postbqppp} immediately
yields that $\mathsf{PP}$\ is closed under intersection, as well as under
$\mathsf{BQP}$\ truth-table reductions.

\section{Fantasy Quantum Mechanics\label{FANTASY}}

Is quantum mechanics an island in theoryspace? \ By \textquotedblleft
theoryspace,\textquotedblright\ I mean the space of logically conceivable
physical theories, with two theories close to each other if they differ in few
respects. \ An \textquotedblleft island\textquotedblright\ in theoryspace is
then a natural and interesting theory, whose neighbors are all somehow
perverse or degenerate. \ The Standard Model is not an island, because we do
not know of any compelling (non-anthropic) reason why the masses and coupling
constants should have the values they do. \ Likewise, general relativity is
probably not an island, because of alternatives such as the Brans-Dicke theory.

To many physicists, however, quantum mechanics \textit{does} seem like an
island: change any one aspect, and the whole theory becomes inconsistent or
nonsensical. \ There are many mathematical results supporting this opinion:
for example, Gleason's Theorem \cite{gleason}\ and other \textquotedblleft
derivations\textquotedblright\ of the $\left\vert \psi\right\vert ^{2}%
$\ probability rule \cite{deutsch:dec,zurek}; arguments for why amplitudes
should be complex numbers, as opposed to (say) real numbers or quaternions
\cite{aar:isl,cfs,hardy}; and \textquotedblleft absurd\textquotedblright%
\ consequences of allowing nonlinear transformations between states
\cite{al,gisin,polchinski}. \ The point of these results is to provide some
sort of explanation for why quantum mechanics has the properties it does.

In 1998, Abrams and Lloyd \cite{al}\ suggested that computational complexity
could also be pressed into such an explanatory role. \ In particular, they
showed that under almost any nonlinear variant of quantum mechanics, one could
build a \textquotedblleft nonlinear quantum computer\textquotedblright\ able
to solve $\mathsf{NP}$-complete and even $\mathsf{\#P}$-complete problems in
polynomial time.\footnote{A caveat is that it remains an open problem whether
this can be done fault-tolerantly. \ The answer might depend on the allowed
types of nonlinear gate. \ On the other hand, if arbitrary $1$-qubit nonlinear
gates can be implemented without error, then even $\mathsf{PSPACE}$-complete
problems can be solved in polynomial time. \ This is tight, since nonlinear
quantum computers are not hard to simulate in $\mathsf{PSPACE}$.} \ One
interpretation of their result is that we should look very hard for
nonlinearities in experiments! \ But a different interpretation, the one I
prefer, is that their result provides independent evidence that quantum
mechanics is linear.

This section builds on Theorem \ref{postbqppp}\ to offer similar
\textquotedblleft evidence\textquotedblright\ that quantum mechanics is
unitary, and that the measurement rule is $\left\vert \psi\right\vert ^{2}$.

Let $\mathsf{BQP}_{\text{\textsf{nu}}}$\ be the class of problems solvable by
a uniform family of polynomial-size, bounded-error quantum circuits, where the
circuits can consist of arbitrary $1$- and $2$-qubit \textit{invertible}
linear transformations, rather than just unitary transformations.
\ Immediately before a measurement, the amplitude $\alpha_{x}$\ of each basis
state $\left\vert x\right\rangle $\ is divided by $\sqrt{\sum_{y}\left\vert
\alpha_{y}\right\vert ^{2}}$ to normalize it.

\begin{proposition}
\label{nuglobal}$\mathsf{BQP}_{\text{\textsf{nu}}}=\mathsf{PP}$.
\end{proposition}

\begin{proof}
The inclusion $\mathsf{BQP}_{\text{\textsf{nu}}}\subseteq\mathsf{PP}$\ follows
easily from Adleman, DeMarrais, and Huang's proof that $\mathsf{BQP}%
\subseteq\mathsf{PP}$ \cite{adh}, which does not depend on unitarity. \ For
the other direction, by Theorem \ref{postbqppp}\ it suffices to show that
$\mathsf{PostBQP}\subseteq\mathsf{BQP}_{\text{\textsf{nu}}}$. \ To postselect
on a qubit being $\left\vert 1\right\rangle $, simply apply the $1$-qubit
nonunitary operation%
\[
\left(
\begin{array}
[c]{cc}%
2^{-q\left(  n\right)  } & 0\\
0 & 1
\end{array}
\right)
\]
for some sufficiently large polynomial $q$.
\end{proof}

Next, for any nonnegative real number $p$, define $\mathsf{BQP}_{p}%
$\ similarly to $\mathsf{BQP}$, except that when we measure, the probability
of obtaining a basis state $\left\vert x\right\rangle $ equals $\left\vert
\alpha_{x}\right\vert ^{p}/\sum_{y}\left\vert \alpha_{y}\right\vert ^{p}$
rather than $\left\vert \alpha_{x}\right\vert ^{2}$. \ Thus $\mathsf{BQP}%
_{2}=\mathsf{BQP}$. \ Assume that all gates are unitary and that there are no
intermediate measurements, just a single standard-basis measurement at the end.

\begin{theorem}
\label{bqpp}$\mathsf{PP}\subseteq\mathsf{BQP}_{p}\subseteq\mathsf{PSPACE}$ for
all constants $p\neq2$, with $\mathsf{BQP}_{p}=\mathsf{PP}$\ when
$p\in\left\{  4,6,8,\ldots\right\}  $.
\end{theorem}

\begin{proof}
To simulate $\mathsf{PP}$ in $\mathsf{BQP}_{p}$, run the algorithm of Theorem
\ref{postbqppp},\ having initialized $O\left(  n^{2}q\left(  n\right)
/\left\vert 2-p\right\vert \right)  $\ ancilla qubits to $\left\vert
0\right\rangle $ for some sufficiently large polynomial $q$. \ Suppose the
algorithm's state at some point is $\sum_{z}\alpha_{z}\left\vert
z\right\rangle $, and we want to postselect on the event $\left\vert
z\right\rangle \in\mathcal{S}$, where $\mathcal{S}$ is a subset of basis
states. \ Here is how: if $p<2$, then\ apply Hadamard gates to $K=2q\left(
n\right)  /\left(  2-p\right)  $\ fresh ancilla qubits conditioned on
$\left\vert z\right\rangle \in\mathcal{S}$. \ The result is to increase the
\textquotedblleft probability mass\textquotedblright\ of each $\left\vert
z\right\rangle \in\mathcal{S}$\ from $\left\vert \alpha_{z}\right\vert ^{p}%
$\ to%
\[
2^{K}\cdot\left\vert 2^{-K/2}\alpha_{z}\right\vert ^{p}=2^{\left(  2-p\right)
K/2}\left\vert \alpha_{z}\right\vert ^{p}=2^{q\left(  n\right)  }\left\vert
\alpha_{z}\right\vert ^{p},
\]
while the probability mass of each $\left\vert z\right\rangle \notin
\mathcal{S}$\ remains unchanged. \ Similarly, if $p>2$, then apply Hadamard
gates to $K=2q\left(  n\right)  /\left(  p-2\right)  $\ fresh ancilla qubits
conditioned on $\left\vert z\right\rangle \notin\mathcal{S}$. \ This decreases
the probability mass of each $\left\vert z\right\rangle \notin\mathcal{S}%
$\ from $\left\vert \alpha_{z}\right\vert ^{p}$\ to $2^{K}\cdot\left\vert
2^{-K/2}\alpha_{z}\right\vert ^{p}=2^{-q\left(  n\right)  }\left\vert
\alpha_{z}\right\vert ^{p}$,\ while the probability mass of each $\left\vert
x\right\rangle \in\mathcal{S}$\ remains unchanged. \ The final observation is
that Theorem \ref{postbqppp}\ still goes through if $p\neq2$. \ For it
suffices to distinguish the case$\ \left\vert \left\langle +|\varphi_{2^{i}%
}\right\rangle \right\vert >0.985$ from $\left\vert \left\langle
+|\varphi_{2^{i}}\right\rangle \right\vert \leq1/\sqrt{2}$\ with exponentially
small probability of error, using polynomially many copies of the state
$\left\vert \varphi_{2^{i}}\right\rangle $. \ But we can do this for any
$p$,\ since all $\left\vert \psi\right\vert ^{p}$\ rules behave well under
tensor products (in the sense that $\left\vert \alpha\beta\right\vert
^{p}=\left\vert \alpha\right\vert ^{p}\left\vert \beta\right\vert ^{p}$).

The inclusion $\mathsf{BQP}_{p}\subseteq\mathsf{PSPACE}$\ follows easily from
the techniques used by Bernstein and Vazirani \cite{bv}\ to show
$\mathsf{BQP}\subseteq\mathsf{PSPACE}$. \ Let $\mathcal{S}$ be the set of
accepting states; then simply compute $\sum_{z\in S}\left\vert \alpha
_{z}\right\vert ^{p}$\ and $\sum_{z\notin S}\left\vert \alpha_{z}\right\vert
^{p}$\ and see which is greater.

To simulate $\mathsf{BQP}_{p}$ in $\mathsf{PP}$ when $p\in\left\{
4,6,8,\ldots\right\}  $, we generalize the result of Adleman, DeMarrais, and
Huang \cite{adh}, which handled the case $p=2$. \ As in Proposition
\ref{adhgen}, we can write each amplitude $\alpha_{z}$ as a sum of
exponentially many contributions, $a_{z,1}+\cdots+a_{z,N}$, where each
$a_{z,i}$\ is a rational real number computable in classical polynomial
time.\ \ Then letting $\mathcal{S}$ be the set of accepting states, it
suffices to test whether%
\begin{align*}
\sum_{z\in\mathcal{S}}\left\vert \alpha_{z}\right\vert ^{p} &  =\sum
_{z\in\mathcal{S}}\alpha_{z}^{p}\\
&  =\sum_{z\in\mathcal{S}}\left(  \sum_{i\in\left\{  1,\ldots,N\right\}
}a_{z,i}\right)  ^{p}\\
&  =\sum_{z\in\mathcal{S}}\sum_{B\subseteq\left\{  1,\ldots,N\right\}
,\left\vert B\right\vert =p}%
{\displaystyle\prod\limits_{i\in B}}
a_{z,i}%
\end{align*}
is greater than $\sum_{z\notin\mathcal{S}}\left\vert \alpha_{z}\right\vert
^{p}$. \ We can do this in $\mathsf{PP}$ by separating out the positive and
negative contributions $\prod_{i\in B}a_{z,i}$, exactly as in Proposition
\ref{adhgen}.
\end{proof}

\section{Open Problems\label{OPENPOST}}

What other classical complexity classes can we characterize in quantum terms,
and what other questions can we answer by that means? \ A first step might be
to prove even stronger closure properties for $\mathsf{PP}$. \ For example,
let $\mathsf{BQP}_{\mathsf{\Vert}}^{\mathsf{PostBQP}}$\ be the class of
problems solvable by a $\mathsf{BQP}$\ machine\ that can make a single
\textit{quantum} query, which consists of a list of polynomially many
questions for a $\mathsf{PostBQP}$ oracle. \ Then does $\mathsf{BQP}%
_{\mathsf{\Vert}}^{\mathsf{PostBQP}}$ equal $\mathsf{PostBQP}$? \
The difficulty in showing this seems to be uncomputing garbage
qubits after the $\mathsf{PostBQP}$\ oracle is simulated.

Also, can we use Theorem \ref{postbqppp} to give a simple quantum
proof of Beigel's result \cite{beigel}\ that
$\mathsf{P}^{\mathsf{NP}}\not \subset \mathsf{PP}$\ relative to an
oracle?

As for fantasy quantum mechanics, an interesting open question is whether
$\mathsf{BQP}_{p}=\mathsf{PP}$ for all nonnegative real numbers $p\neq2$. \ A
natural idea for simulating $\mathsf{BQP}_{p}$\ in $\mathsf{PP}$ would be to
use a Taylor series expansion for the probability masses $\left\vert
\alpha_{x}\right\vert ^{p}$. \ Unfortunately, I have no idea how to get fast
enough convergence.

\section{Acknowledgments}

I thank Avi Wigderson for helpful discussions, and Richard Beigel
and Lance Fortnow for correspondence.

\bibliographystyle{plain}
\bibliography{thesis}

\end{document}